\documentclass[nocits]{PoS}
\usepackage{natbib}

\def\aap{\ref@jnl{A\&A}}

\let\ifpdf\relax

\newcommand{\skipthis}[1]{}

\title{Galaxy Formation \& Dark Matter Modelling in the Era of the Square Kilometre Array}

\author{C. Power$^{a\ast}$, C.D.P. Lagos$^{b\dagger}$, B. Qin$^{c\ddagger}$, C.M. Baugh$^d$, D. Cunnama$^e$, J. Fu$^f$, 
  H.S. Kim$^g$, C.G. Lacey$^d$, L. Li$^c$, D. Obreschkow$^a$, J. Wang$^c$, Y. Wang$^c$ \& M. Zhu$^c$\\
  \llap{$^a$} International Centre for Radio Astronomy Research,The University of Western Australia, 35 Stirling Highway,Crawley, WA 6009, Australia\\
  \llap{$^b$} European Southern Observatory,Karl-Schwarzschild-Str. 2, 85748 Garching bei M\"unchen, Germany\\
  \llap{$^c$} National Astronomical Observatories, Chinese Academy of Science, A20 Datun Road, Beijing 100012, China\\
  \llap{$^d$} Institute for Computational Cosmology, Durham University, Science Laboratories, South Road, Durham DH1 3LE, United Kingdom\\
  \llap{$^e$} Department of Physics, University of Western Cape, Bellville 7535, Republic of South Africa\\
  \llap{$^f$} Shanghai Astronomical Observatory, Chinese Academy of Sciences, 80 Nandan Road, Shanghai 200030, China\\
  \llap{$^g$} School of Physics, The University of Melbourne, Parville, Vic 3010, Australia\\
  E-mail: $^{\ast}$\email{chris.power@icrar.org}, $^{\dagger}$\email{clagos@eso.org}, $^{\ddagger}$\email{qinbo@bao.ac.cn}}

\abstract{Theoretical galaxy formation models are an established and 
powerful tool for interpreting the astrophysical significance of observational
data, particularly galaxy surveys. Such models have been utilised with 
great success by optical surveys such as 2dFGRS and SDSS, but their application 
to radio surveys of cold gas in galaxies has been limited. In this chapter we describe 
recent developments in the modelling of the cold gas properties in the models, and how
these developments are essential if they are to be applied to cold gas surveys 
of the kind that will be carried out with the SKA. By linking explicitly 
a galaxy's star formation rate to the abundance of molecular hydrogen in the 
galaxy rather than cold gas abundance, as was assumed previously, 
the latest models reproduce naturally many of the global atomic and molecular 
hydrogen properties of observed galaxies. We review some of the key results of 
the latest models and highlight areas where further developments are 
necessary. We discuss also how model predictions can be most 
accurately compared with observational data, what challenges we expect when 
creating synthetic galaxy surveys in the SKA era, and how the SKA can be used to test
models of dark matter.}

\FullConference{
Advancing Astrophysics with the Square Kilometre Array\\
June 8-13, 2014\\
Giardini Naxos, Italy}

\ShortTitle{Galaxy Formation, Dark Matter \& the SKA}

\begin{document}

\section{Introduction}
\label{Sec:Intro}

Neutral hydrogen, both in its atomic (HI) and molecular (H$_2$) forms, plays a 
fundamental role in galaxy formation, principally as the raw material from which 
stars are made. At any given time, the fraction of a galaxy's mass that is in 
the form of neutral hydrogen will be determined by the competing rates at which 
it is depleted (by e.g. star formation, photo-ionization, expulsion via winds) 
and replenished (by e.g. recombination, accretion from galaxy's surroundings) -- 
physical processes that are the foundations of any viable theory of galaxy formation. 

Most of what we know currently about HI in galaxies derives from 21-cm surveys of 
the nearby Universe ($z \lesssim 0.05$) such as HIPASS 
\citep[HI Parkes All Sky Survey; cf.][]{Barnes01} and ALFALFA 
\citep[Arecibo Legacy Fast ALFA; cf.][]{Giovanelli05}. Efforts are ongoing 
to complement these data with results from higher redshifts, using innovative techniques 
to estimate statistically typical galaxy HI content (e.g. stacking, cf. \citealt{Lah07} and 
\citealt{Delhaize13}). These efforts will accelerate dramatically over the coming decade, 
with the advent of next generation radio telescopes such as the Australian Square Kilometre 
Array Pathfinder \citep[ASKAP][]{Johnston08}, MeerKAT \citep{DeBlok09}, and ultimately the 
Square Kilometre Array (SKA) itself. The SKA and its {shorter} wavelength counterpart the 
Atacama Large Millimetre Array (hereafter ALMA) offer the technical capability to chart the 
evolution of neutral hydrogen in galaxies over the last $\sim $12 billion years of 
cosmic time. This is a significant fraction of the assembly history of present-day 
galaxies, and the results of these observations will provide a stringent test of our 
theories of galaxy formation and evolution. 

These theories of galaxy formation and evolution have undergone 
some timely and fundamental changes over the last $\sim$5 years. Signficantly for 
forthcoming HI and H$_2$ galaxy surveys, there has been particular emphasis 
on model predictions for the cold gas content of galaxies, which has led to some 
important conceptual changes in, for example, the treatment of star formation in 
the models. In the remainder of this chapter, we review briefly the essential 
features of the theoretical galaxy formation models, in particular the semi-analytical 
approach (\S\ref{sec:overview}); discuss the recent developments in these models, explaining 
why they were necessary (\S\ref{sec:sams}) and what the consequences are (\S\ref{sec:sf} to 
\S\ref{sec:kinematics}); and indicate where further 
developments will be necessary (\S\ref{Future}). Finally, we explore the potential of the 
SKA as a testbed for our theories of dark matter (\S\ref{sec:darkmatter}).

\section{Galaxy Formation Modelling: Hydrodynamical and Semi-Analytical Approaches}
\label{sec:overview}

\begin{figure}[t]
\centering
\includegraphics[width=0.55\textwidth]{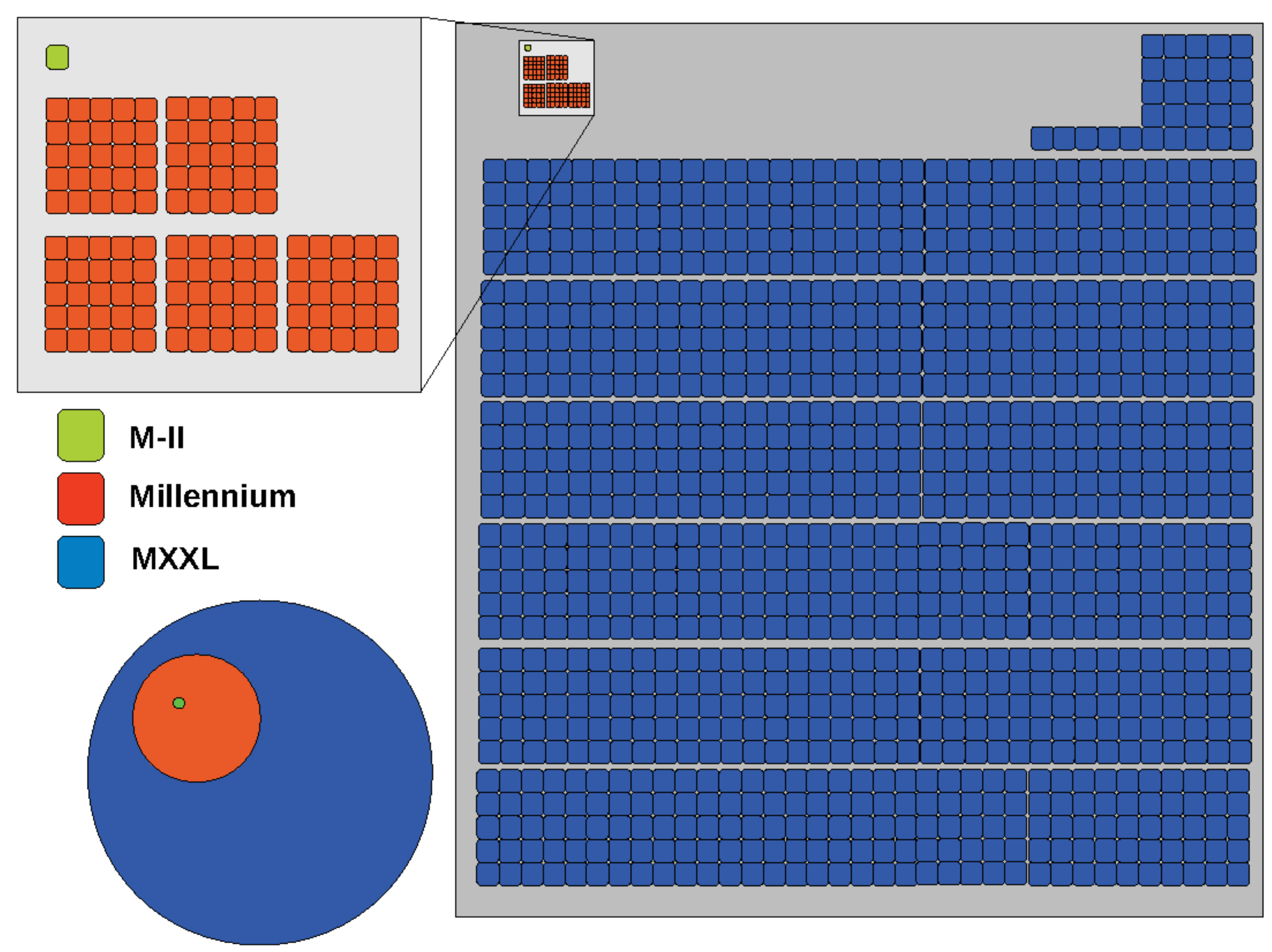} 
\caption{Schematic figure showing the difference in volume between the 
Millennium-I, II and XXL simulations (courtesy of Raul Angulo). The volume of 
Millennium-I exceeds that of Millennium-II by a factor of $125$ times, while it is 
dwarfed by XXL by more than a factor of $200$ in volume. The particle mass of 
Millennium-II is $\approx 1225$ times smaller than the XXL simulation. Galaxy formation
models for the SKA will require Millennium-II resolution in Millennium-I or XXL volumes.} 
\label{MillComp}
\end{figure}

The two most widely used techniques to study galaxy formation in a cosmological context
are \emph{hydrodynamical $N$-body simulations} and \emph{semi-analytical models}. Both follow
the formation and evolution of dark matter halos over cosmic time using cosmological $N$-body
simulations. However, hydrodynamical models compute the formation and evolution of the baryonic
structures (e.g. hot gaseous halos, galaxy discs) at the same time as the dark matter, whereas
semi-analytical models use the results of the pre-computed $N$-body simulation to guide the 
formation and evolution of the baryonic structures subsequently. 

The methods should be viewed as complementary, each with their own advantages and disadvantages. 
Briefly, these are as follows.

\begin{itemize}

\item \emph{Hydrodynamical $N$-body simulations} capture the complexity of gas and stellar 
dynamics in a way that semi-analytical models cannot, which is invaluable when studying, for 
example, the details of gas accretion onto galaxies. They are, however, computationally expensive 
when compared to semi-analytical models, and usually employed when studying galaxy formation 
on the scale of individual systems or in small-to-intermediate cosmological volumes. 

\item \emph{Semi-analytical models} are computationally inexpensive when 
compared to hydrodynamical simulations, requiring only the $N$-body simulation and derived 
merger histories for the host dark matter halos of galaxies. While they sacrifice capturing 
the complex dynamical and structural properties of galaxies, they are ideally suited to the 
study of statistical samples of galaxies in cosmological volumes over many cosmic epochs. 
The models are such that they can be parameterised and calibrated to reproduce a set of standard 
observations (e.g. the galaxy luminosity function, and the Tully-Fisher relation) in a
relatively straightforward manner (see \citealt{Baugh06} for a comprehensive review), a process
that is comparatively unwieldy for hydrodynamical simulations. 

\end{itemize}

\noindent Therefore, it is in the detailed dynamical and structural properties of baryons 
within galaxies and in the surrounding haloes that the methods differ\footnote{Although we note 
  that efforts are underway in the semi-analytical models to capture, for example, the effect 
  of baryon displacement highlighted by e.g. \citet{vandaalen.etal.2014} in hydrodynamical
  simulations}; otherwise both approaches treat the physics of galaxy formation 
-- star formation, stellar winds and supernovae, black hole growth and outburst of feedback -- 
using similar simplified prescriptions that parameterise physical processes on spatial scales 
much smaller than can be resolved in the simulations.

This distinction is important when considering the kinds of volumes ($1\, {\rm Gpc}^3$ and 
larger) that the SKA will be probing. To date, the state-of-the-art in cosmological 
hydro-dynamical galaxy formation simulations probe volumes of order $100~{\rm Mpc}^3$ 
(\citealt{Vogelsberger14}; \citealt{Schaye14}). Contrast with semi-analytical models, which 
have been applied to the Millennium series of simulations. These are some of the largest 
cosmological $N$-body (i.e. dark matter only) simulations ever run, both in terms of particle 
number and volume -- spanning the Millennium-I Simulation (\citealt{Springel05}; with particle 
mass of $1.2\times 10^{9}\,h^{-1} \rm M_{\odot}$ and box size of $500 h^{-1} \rm Mpc$ ), the 
Millennium-II Simulation (\citealt{Boylan-Kolchin09}; with particle mass of 
$6.9\times 10^{6}\,h^{-1} \rm M_{\odot}$ and box size of $100 h^{-1} \rm Mpc$) and 
Millennium XXL (\citealt{Angulo12}; with particle mass of 
$8.5\times 10^{9}\,h^{-1} \rm M_{\odot}$ and box size of $3000 h^{-1} \rm Mpc$). 
Fig.~\ref{MillComp} shows these three simulations in a way that emphasises the differences
in volumes modelled.

Fig.~\ref{MillComp} also encapsulates the challenge facing galaxy formation modelling in the SKA era.
The SKA will probe enormous volumes, in excess of those modelled in the Millennium-I
and comparable to those in Millennium XXL, but most HI in the Universe is concentrated in low-mass
galaxies and dark matter halos; this requires high mass resolution, comparable to that of the
Millennium-II. This combination of high mass resolution and enormous volume demands
a semi-analytical rather than a hydrodynamical approach. Hydrodynamical simulations will play a 
vital role in informing semi-analytic models, both about the limitations of assumptions as well 
as improvements and approximations that can be made (some aspects of which we discuss in 
Section~\ref{Future}). However, it is semi-analytical models that we expect to underpin the 
theoretical framework supporting the SKA and it is this approach that we focus on in the 
remainder of this chapter. 

\section{Semi-Analytical Modelling of HI in Galaxies}
\label{sec:sams}

As noted already, semi-analytical models offer a simplified physically motivated 
treatment of the processes that control the amount of cold gas\footnote{Here cold gas refers 
to gas that has cooled radiatively from a hot phase to below $10^{4}\rm K$, is predominately HI,
H$_2$ and helium (He), and is available for star formation.} in a galaxy: namely the rate at 
which it cools radiatively within its host dark matter halo; the rate at which it is accreted in 
galaxy mergers; the rate at which it is consumed in star formation; and the rate at which it is 
reheated or expelled from galaxies by sources of feedback (e.g. photo-ionisation, stellar winds, 
supernovae and active galactic nuclei, AGN, heating).

\begin{figure}[t]
\centering
\includegraphics[width=\textwidth]{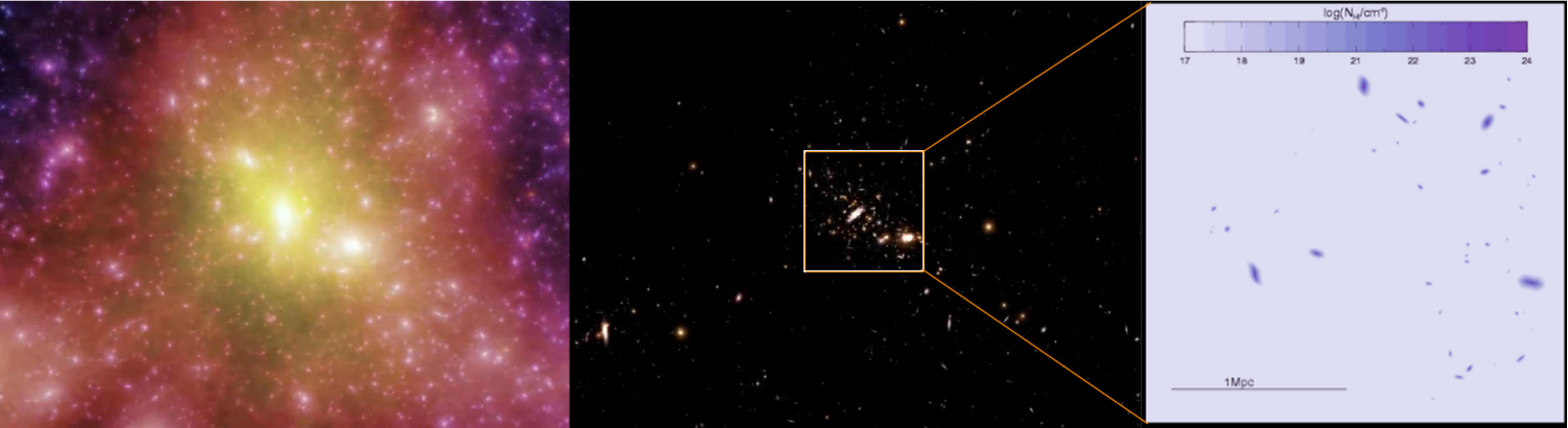} 
\caption{A massive galaxy cluster drawn from the Millennium-I Simulation, showing colour coded 
  projected dark matter density, galaxy optical light and HI content (left, middle and right panels 
  respectively). Galaxy properties were calculated using the \small{GALFORM} semi-analytical model 
  of \citet{Lagos11}. Image of dark matter distribution courtesy of Volker Springel.} 
\label{fig:hi_gals_dm}
\end{figure}

\begin{figure}[h]
\centering
\includegraphics[width=0.43\textwidth]{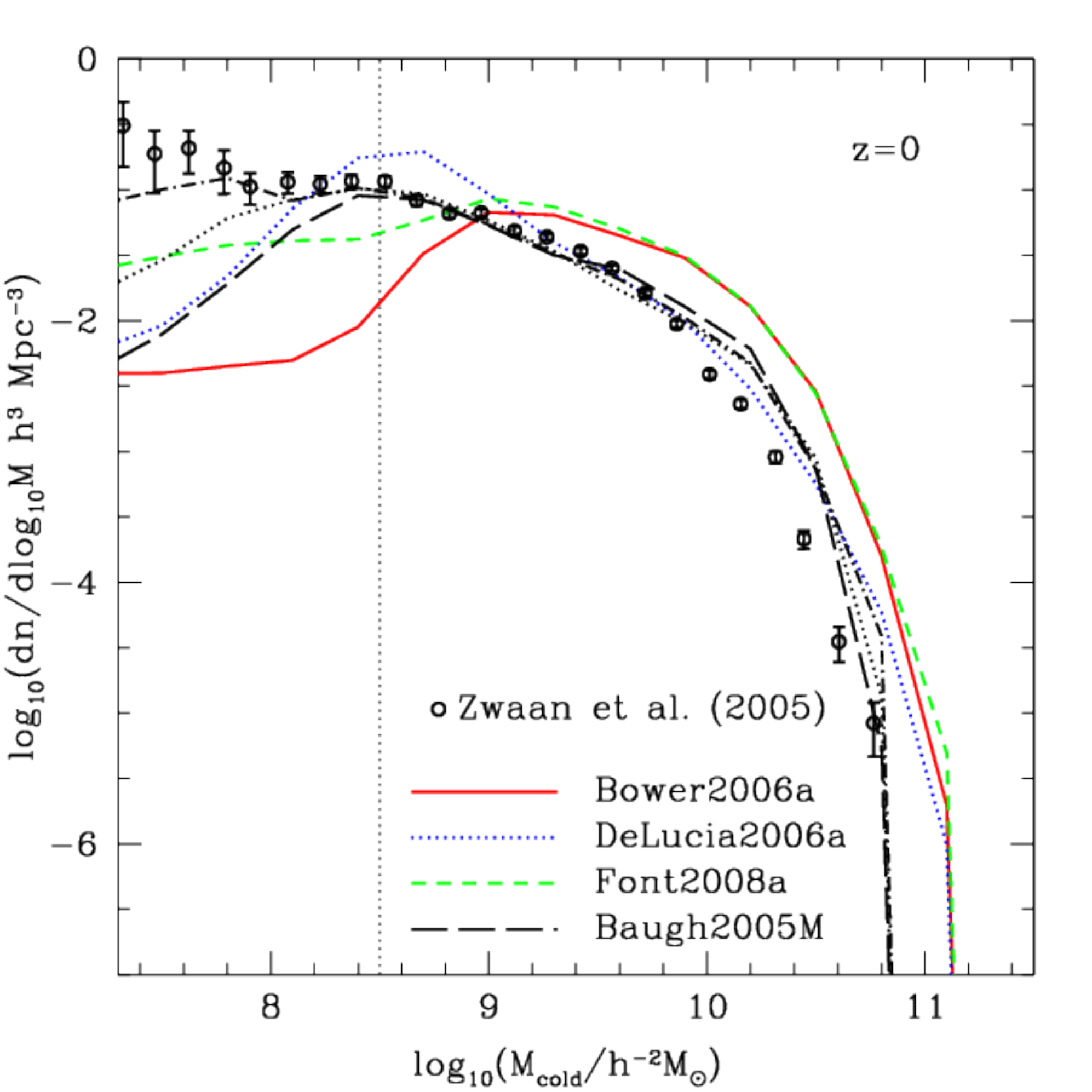} 
\includegraphics[width=0.43\textwidth]{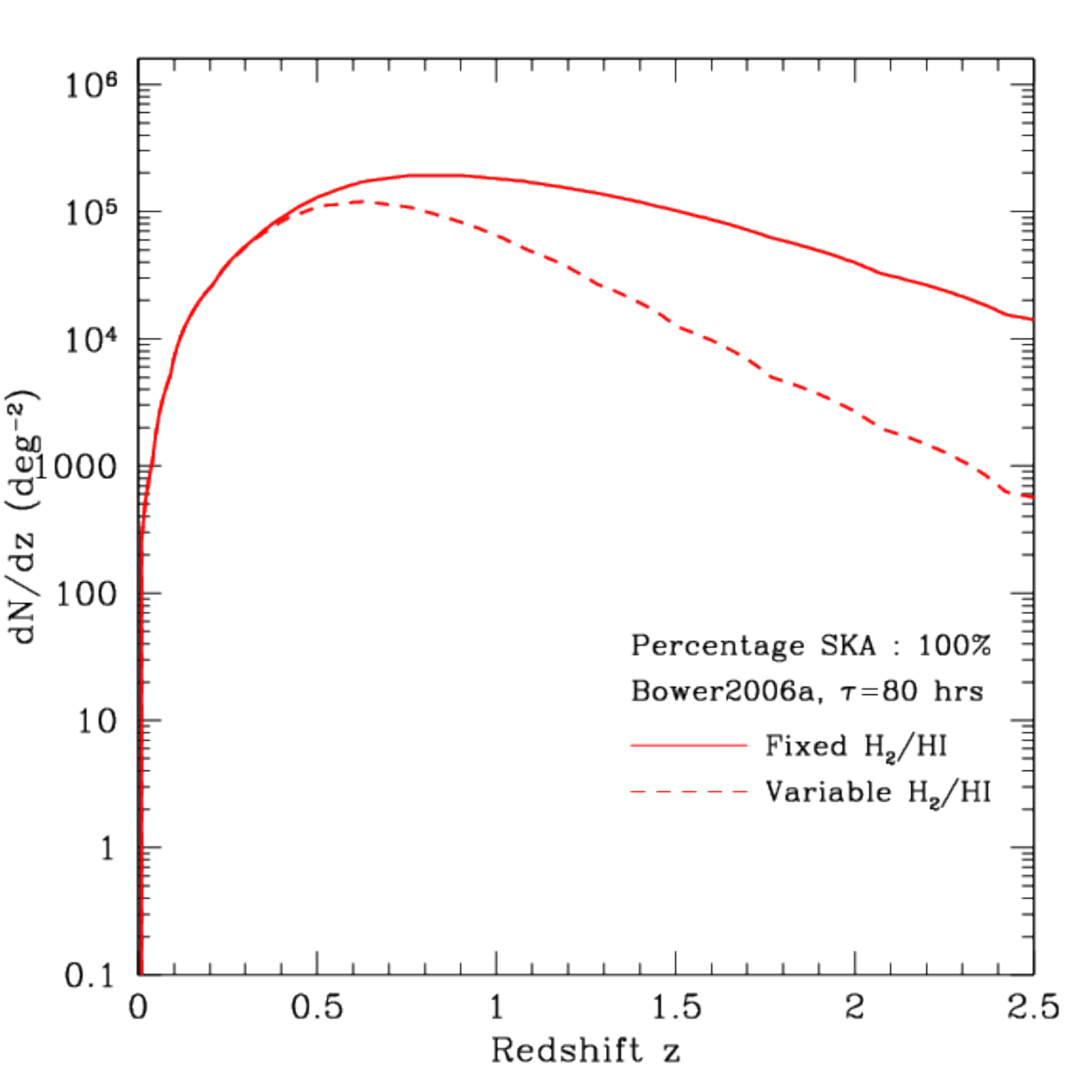} 
\caption{\emph{Left Panel:} Cold gas mass function at $z=0$ in 4 different galaxy formation 
  models applied to the Millennium-I Simulation (see legend); data points are observational 
  estimates derived from HIPASS \citep{Zwaan05} obtained by rescaling HI masses by a factor 
  of $\sim$1.85; see text for further details. Dotted vertical line corresponds to the resolution 
  limit of the models. \emph{Right Panel:} Predicted number counts of 
  galaxies per square degree per unit redshift for peak flux limited 1-year survey on a fiducial 
  SKA with effective area 1 $\rm km^2$ and integration time of $\tau$=80 hrs, based 
  on \citet{Bower06}. Solid (dashed) curves correspond to a fixed (variable) conversion factor 
  from cold gas to $\rm H_2/HI$. Both figures from \citet{Power10}.} 
\label{fig:mass_function}
\end{figure}

A visual impression of the results of a semi-analytical calculation is given in 
Fig.~\ref{fig:hi_gals_dm}, which shows the stellar and HI content of galaxies in a 
cluster drawn from the Millennium-I Simulation, as predicted by the \citet{Lagos11} model. 
The \citet{Lagos11} model was one of the first to predict the HI and H$_2$ content of galaxies 
explicitly and self-consistently, tracking individual gas phases over the duration of 
the calculation; prior to this, HI and H$_2$ masses in galaxies were 
obtained by post-processing cold gas masses at the end 
of the calculation. We consider briefly the results of these older \emph{post-processed} models, 
and the physics underpinning the newer \emph{self-consistent} models.

\emph{Post-processed models} predicted galaxy populations with cold gas properties in 
broad agreement with observations. \citet{Baugh04} and \citet{Rawlings04} showed that 
semi-analytical mass functions of HI in galaxies (hereafter HIMF) were in satisfactory agreement
with observations \citep[HIPASS; cf.][]{Zwaan03} and semi-empirical fits. They converted 
between cold gas mass and HI mass by using the observationally measured global densities
of HI and H$_2$ (\citealt{Zwaan03} and \citealt{Keres03} respectively) and splitting the cold
gas mass into HI, H$_2$ and He phases. \citet{Obreschkow09b} and \citet{Obreschkow09} adopted
a more sophisticated approach to conversion, using the empirical relation between the HI/H$_2$ 
ratio and midplane pressure in galaxies \citep[cf.][]{Blitz06} to predict the HI, H$_2$ and He
phases on a galaxy-by-galaxy basis. \citet{Power10} compared results from four then favoured 
models \citep{Baugh05,Bower06,DeLucia07,Font08}
applied to the Millennium-I Simulation for the mass function (MF) of cold neutral gas (atomic and 
molecular) in galaxies as a function of redshift. Despite different implementations of galaxy 
formation physics, the predictions were found to be broadly consistent with one another; key 
differences reflected how the models treated AGN feedback and the timescale for star formation 
as a function of redshift. The predicted MF of cold gas in galaxies agreed well with that 
inferred from HI surveys at $z$=0 (see left panel of Fig.~\ref{fig:mass_function})\footnote{Here 
  the HIPASS HIMF of \citealt{Zwaan05} was converted to a cold gas MF by 
  an empirically-motivated rescaling of HI masses by a factor of $\sim$1.85. This accounts for 
  the 24\% helium and assumes that HI and H$_2$ are split such that the global HI and H$_2$ 
  densities are consistent with the observational estimates of \citet{Zwaan03} and 
  \citet{Keres03}.}. 

However, consideration of how the HIMF evolves with cosmic time and 
what this implies for future HI surveys highlighted that uncertain conversions from 
cold gas mass to HI and H$_2$ masses translate into large variations in model predictions.
The right panel of 
Fig.~\ref{fig:mass_function} shows how predicted HI number counts are 
influenced by the choice of conversion factor -- the solid (dashed) curve assumes 
\citet{Baugh04} (\citealt{Obreschkow09}) approaches. Number counts differ negligibly 
for $z \lesssim 1$, but grow to a factor of $\sim$10 by $z \sim 2.5$, reflecting the increasing 
proportion of cold gas in the form of H$_2$ with increasing $z$, as galaxies become more gas-rich and 
more compact.

\begin{figure}[t]
\centering
\includegraphics[width=0.6\textwidth]{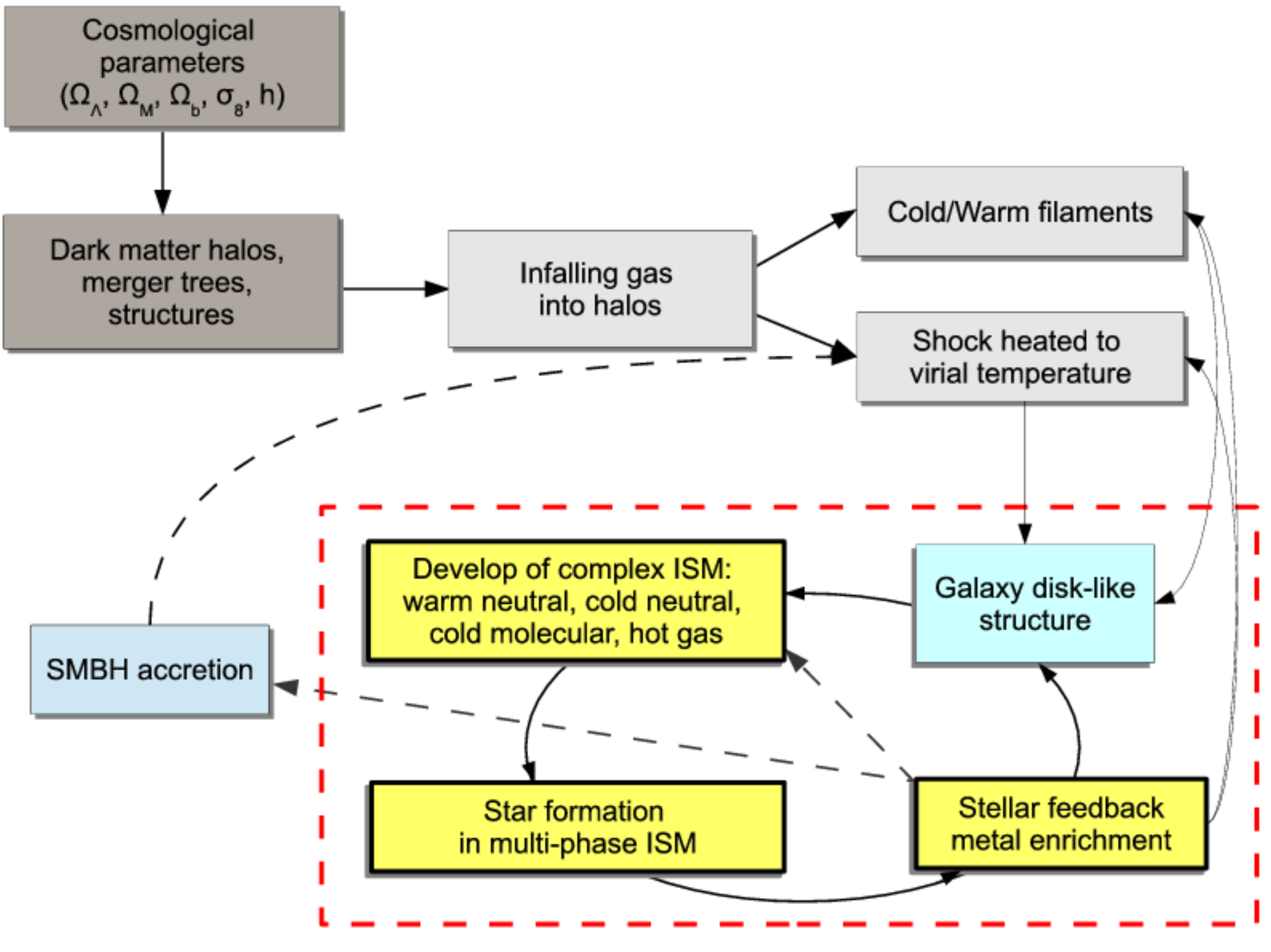} 
\caption{{\bf Schematic of the Physics of Disc Galaxy Formation and the ISM.} Galaxies form within 
the potential wells of dark matter halos, whose formation is governed by the cosmological 
parameters and dark matter model, from baryons; these baryons either shock-heated 
upon infall and cooled or fell directed onto the galaxy along filaments. This emphasises
the impact modelling of the ISM and SF has on galaxy components and the circum- and 
inter-galactic medium.}
\label{fig:Chart_ISM}
\end{figure}

\emph{Self-consistent models} have addressed the question of how to convert from a cold gas mass to 
HI and H$_2$ mass by focusing on improved modelling of the interstellar medium (ISM), and, in 
particular, the relationship between star formation (SF) and neutral hydrogen phases; this has 
been the pivotal area of development in semi-analytical models over the last $\sim$5 years. 
Post-processed models treated the ISM as a single star-forming phase (e.g. \citealt{Cole00}),
but this is at odds with both empirical evidence and theoretical expectation. Observationally, 
there is evidence that a galaxy's SF rate (SFR) per unit area and molecular gas surface 
density correlate linearly, $\Sigma_{\rm SFR}\propto\Sigma_{\rm mol}$ (\citealt{Wong02,Bigiel08}). 
Theoretically, recent hydrodynamical simulations of star forming clouds suggests that the physical 
conditions necessary for SF -- high density and low temperature gas -- are also necessary for 
formation of H$_2$ and CO, but these conditions are driven by the presence of dust shielding, which 
prevents photo-dissociation by energetic UV photons, and CII cooling \citep{Glover12a,Glover12}. In 
other words, the presence of H$_2$ correlates naturally with SF (see also the arguments of 
\citealt{Schaye04} and recent numerical work by \citealt{Gnedin11,Feldmann11,Glover12,Krumholz13}).

This has led to the development of semi-analytical models that implement a fully self-consistent 
treatment of the ISM and SF (cf. \citealt{Cook10,Fu10,Lagos10}; also \citealt{Robertson08,Dutton09}).
Fig.~\ref{fig:Chart_ISM} encapsulates the complex interplay of physical processes in the ISM and 
their influence on galaxy formation, ranging from star formation in high density regions 
(e.g. \citealt{Bigiel08}), to supernovae-driven turbulence (e.g. \citealt{Dobbs11}), to 
chemical enrichment of the inter-galactic medium via outflows (e.g. \citealt{Putman12}), and it 
provides a template for the treatment of the ISM in the models. 

Two of the first models to implement this self-consistent approach -- \citet{Lagos10} and 
\citet{Fu10} -- each explored two approaches to estimating the HI and H$_2$ content of the 
ISM\footnote{The \citet{Lagos10} and \citet{Fu10} follow similar prescriptions, the key 
  difference being that \citet{Fu10} follow the radial gas profile of the galaxy disc 
  explicitly.}. The first was the \emph{empirical relation} of \citet{Blitz06}, which relates the 
molecular-to-atomic surface density ratio to hydrostatic pressure within the disc, 
estimating the SFR from the molecular gas surface density using the well measured molecular 
depletion timescale \citet{Bigiel08} (also used by \citealt{Cook10}). The second was the 
\emph{theoretical relation} of \citet{Krumholz09}, which models SF as taking place in turbulent, 
marginally stable clouds, estimating the molecular abundance from the balance between the 
dissociating radiation flux and the formation of molecules on the surface of dust grains. 
Both \citet{Lagos10} and \citet{Fu10} favoured the empirical relation of \citet{Blitz06}, which 
provides a better fit to the HIMF 
at $z=0$ (\citealt{Zwaan05,Martin10}) and predicts a clustering of HI selected galaxies in good 
agreement with observations (cf. \citealt{Kim13}); in contrast, the theoretical model of 
\citet{Krumholz09} overpredicts the number density of intermediate HI mass galaxies. 

In the remainder of this chapter, we assume self-consistent models that derive from 
\citet{Lagos10} and later. Model parameters are calibrated to reproduce the $z$=0 K- and 
b$_j$ band luminosity functions (hereafter LFs), which govern AGN and stellar feedback, 
and UV LFs at $z$=2-6, which limit the duration of starbursts; no gas properties were used.

\section{Insights into Global Star Formation Rates over Cosmic Time}
\label{sec:sf}

The more realistic treatment of the ISM in the newer self-consistent models leads naturally to 
predictions that can explain trends in observational data.

\begin{figure}[]
\includegraphics[width=1\textwidth]{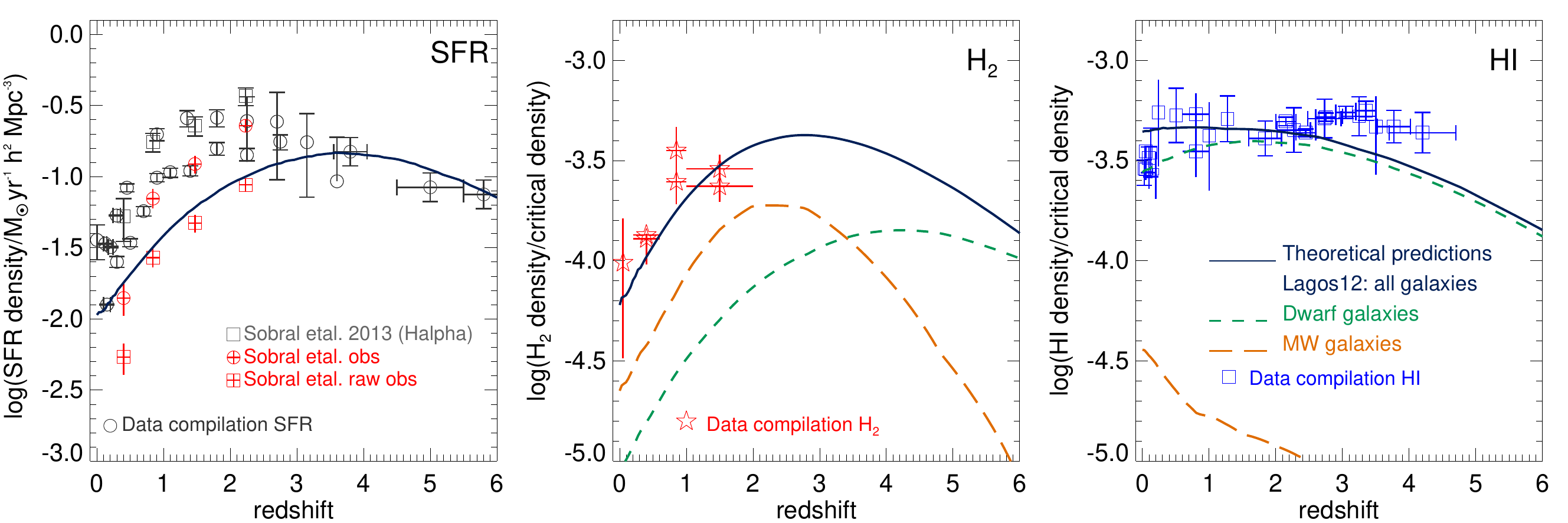}
\caption{Global densities (solid lines) of the SFR in units of $M_{\odot}\, {\rm yr}^{-1}\, h^3\,{\rm Mpc}^{-3}$ 
  (left hand panel), the estimated molecular hydrogen (middle panel) and the atomic hydrogen (right hand panel) in units 
  of the critical density, as a function of redshift for the \citet{Lagos12} model. 
  The circles, stars and squares correspond to a data compilation of observations presented in 
  \citet{Lagos14} for the SFR, H$_2$ and HI densities, and are taken from \citet{Sobral13}. We also show the predicted contribution from dwarf galaxies 
(those with stellar masses $<10^9\,M_{\odot}$; dashed lines) 
and Milky-Way (MW) type galaxies (those with stellar masses in the range 
$10^{10}\,M_{\odot}-10^{11}\,M_{\odot}$; long-dashed lines).}
\label{fig:DensEvo}
\end{figure}

{\emph{First}}, the observed steep decline of SFR density with decreasing redshift is 
a consequence of the steep decline in the underlying molecular gas mass density with redshift 
(see left and middle panels of Fig.~\ref{fig:DensEvo}; see also \citealt{Lagos11}); in contrast, 
HI mass density evolves very weakly with redshift (right panel of Fig.~\ref{fig:DensEvo}). This 
arises from a combination of decreasing gas fractions and increasing galaxy sizes with decreasing 
redshift, which both act to reduce the gas surface density and the hydrostatic pressure of the 
disc. The redshift evolution of SFR density tracks that of the gas surface density of galaxies 
dominating the SFR in the Universe at a given epoch. 

Increasing hydrostatic pressure in galaxy discs and increasing gas fractions 
with increasing redshift also leads to increasing molecular to dynamical mass 
ratios, which is consistent with observations of normal galaxies in the Universe up to 
$z\sim 2.5$ (see \citealt{Geach11}). Note, however, that the increase in the SFR density 
is $2-3$~times higher than that of the H$_2$ density; this reflects the greater efficiency
of star formation in starbursts, which is not accounted for in the normal star-forming discs 
assumed in the model (\citealt{Lagos14}). \citet{Lagos11} showed that the shape of the relation 
between molecular gas fraction (i.e. $M_{\rm mol}/M_{\rm mol}+M_{\ast}$) and redshift depends   
on galaxy environment, such that, on average, galaxies with low mass host halos have larger 
molecular gas fractions than those with more massive hosts. Gas dynamical simulations agree 
qualitatively with this prediction (e.g. \citealt{Dave11}).

The offset between the observational data points and model prediction evident in
the left panel of Fig.~\ref{fig:DensEvo} is well known in galaxy formation models, which tend 
to under-predict the SFR density by factors of $\sim$2-3. This discrepancy has been explored in
some depth in \citet{Lagos14} and is driven by a number of factors, including extrapolation of 
observed luminosity functions to account for the contribution of faint, undetected galaxies; the 
adopted stellar initial MF; and uncertainties in the SFRs of individual galaxies, 
which are of order 0.3-0.4 dex, as found by \citet{Speagle14}.

{\emph{Second}}, the molecular-to-atomic gas ratio correlates with stellar mass, and so the 
low-mass end of the HIMF is steeper than the low-mass end of the stellar and 
H$_2$ MFs. We show in Fig.~\ref{fig:DensEvo} the predicted contribution to the 
HI and H$_2$ densities from low mass galaxies (dwarfs) and more massive, Milky-Way like 
galaxies; the contribution from Milky-Way like galaxies to HI density is negligible except 
at very low redshifts, while dwarf galaxies dominate over the entire redshift range. 
For H$_2$, we see that Milky Way like galaxies dominate at lower redshifts but dwarfs make 
a significant contribution at higher redshifts, when the contribution from 
Milky Way like galaxies declines sharply. Note that these estimates are derived from the 
observations of \citet{Keres03} and \citet{Berta13} and required a number of assumptions to 
be made in their estimation (e.g. correlation between $A_{\rm UV}$ and H$_2$).
Fig.~\ref{fig:DensEvo} suggests that the SKA will provide new and interesting insights into
dwarf galaxies, which at present are only partially accessible optical and near infrared 
telescopes.

\section{Probing the Physics of Feedback with HI Surveys}
\label{sec:feedback}

\begin{figure*}
 \includegraphics[width=1.0\textwidth]{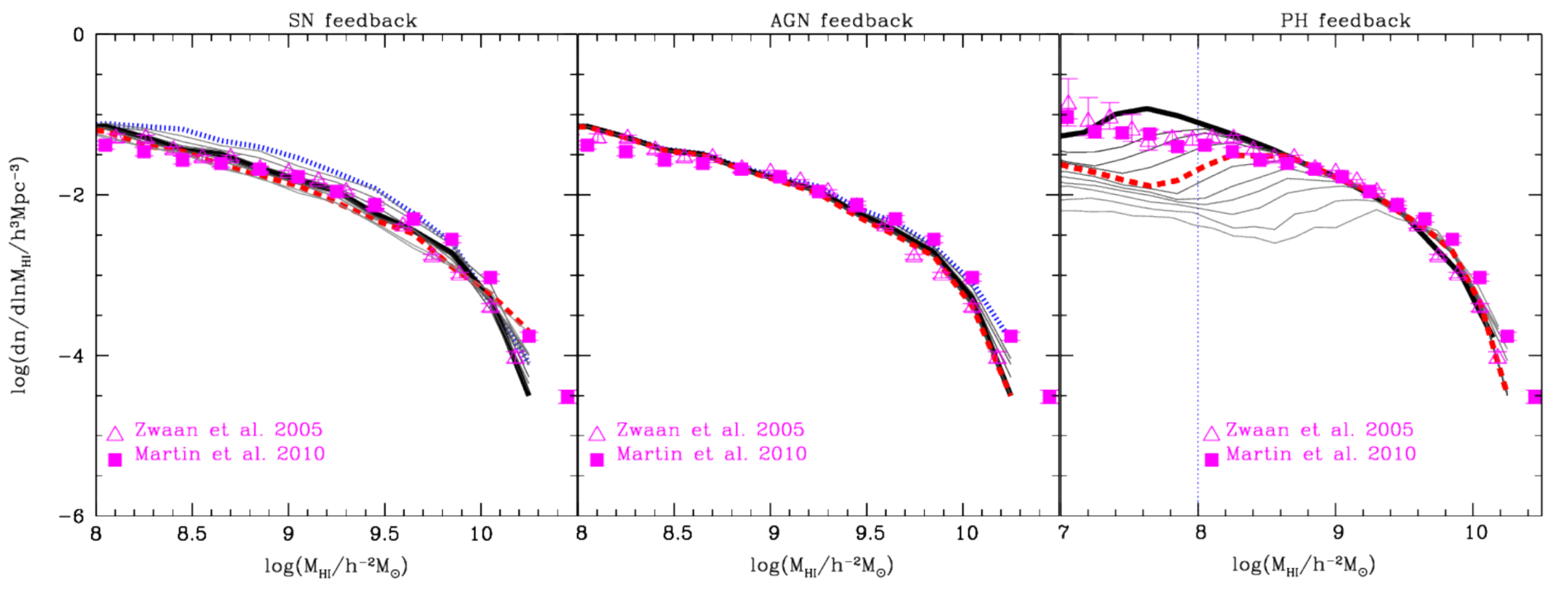}
 \caption{Impact of feedback from supernovae (SNe), AGN and the photoionising 
   background (PH; left to right panels) on the HIMF in the \citet{Lagos11} model. 
   Open triangles (filled squares) represent observational data from the HIPASS \citet{Zwaan05}) 
   and ALFALFA \citet{Martin10} surveys. SNe strength varies in the range of 
   $V_{\rm hot}=\rm 300 km/s$ to $V_{\rm hot}=700\rm km/s$; AGN strength varies in 
   the range of $\alpha_{\rm cool}=0.4-0.8$; photoionisation strength in the range of 
   $30{\rm km/s}\le V_{\rm cut} \le 90 \rm km/s$. Dark solid lines correspond to the 
   default models values; red dashed and blue dotted curves correspond strong and weak 
   feedback limits, respectively.}
  \label{FEED}
\end{figure*}

Semi-analytical models include a range of physical processes that either prevent hot gas 
from cooling onto a galaxy or heat up neutral gas and expel it from a galaxy 
\citep[cf.][]{Baugh06}. Simple physical arguments to detailed numerical calculations 
demonstrate that supernovae are responsible for expelling large amounts of gas from 
galaxies in low mass halos ($M_{\rm halo}<{\rm few} \times 10^{11}\,M_{\odot}$), 
shaping the faint-end of the LF. Similarly, the photo-ionising background 
present during cosmological reionization quenches low-mass galaxy formation, influencing 
the faint-end of the LF \citep[e.g][]{Benson02}. AGN appear necessary to suppress excessive
star formation in high stellar mass galaxies and so regulate the amplitude and shape
of the bright-end of the LF \citep[cf.][]{Bower06,Croton06}.
 
The effect of different feedback mechanisms on optical and near infrared LFs have been explored
with semi-analytical models \citet{Benson03,Baugh05,Bower06,Croton06}, and this has been extended to
the HIMF by \citet{Kim13} using the \citet{Lagos11} model. These authors studied the
influence of varying the strength of supernovae (SNe), AGN and photo-ionizing (PH) feedback 
on both the (optical) b$_j$-band LF and the HIMF; we focus on
the HIMF in Fig.~\ref{FEED}. As the Figure reveals, SNe feedback (left panel) affects
the HIMF uniformly across the entire range of HI masses; weaker feedback boosts the amplitude of
the HIMF, and vice versa. AGN (middle panel) have little impact on the HIMF except at the very 
highest HI masses, with weaker feedback reducing the slope. Interestingly, photo-ionization 
(PH; right panel) has a significant effect at lower HI masses, driven by the sensitivity of the 
low-mass end slope to the halo mass scale below which galaxy formation is suppressed at reionization.
Here the caveat is that particular models for feedback were assumed.

\section{Clustering of HI-Selected Galaxies}
\label{sec:clustering}

\begin{figure}[h]
\centering
 \includegraphics[width=0.4\textwidth]{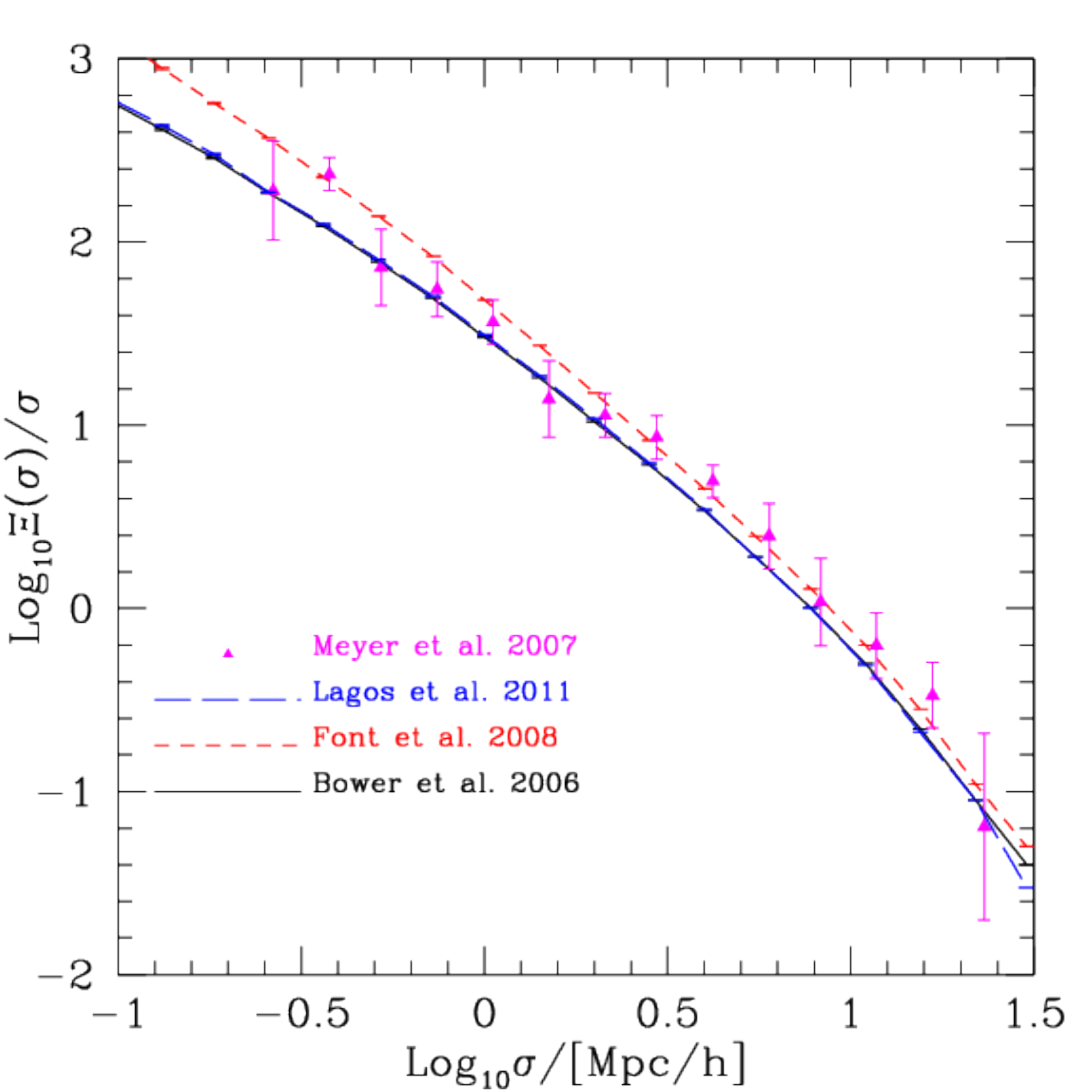}
 \includegraphics[width=0.43\textwidth,trim=0 147 0 0, clip]{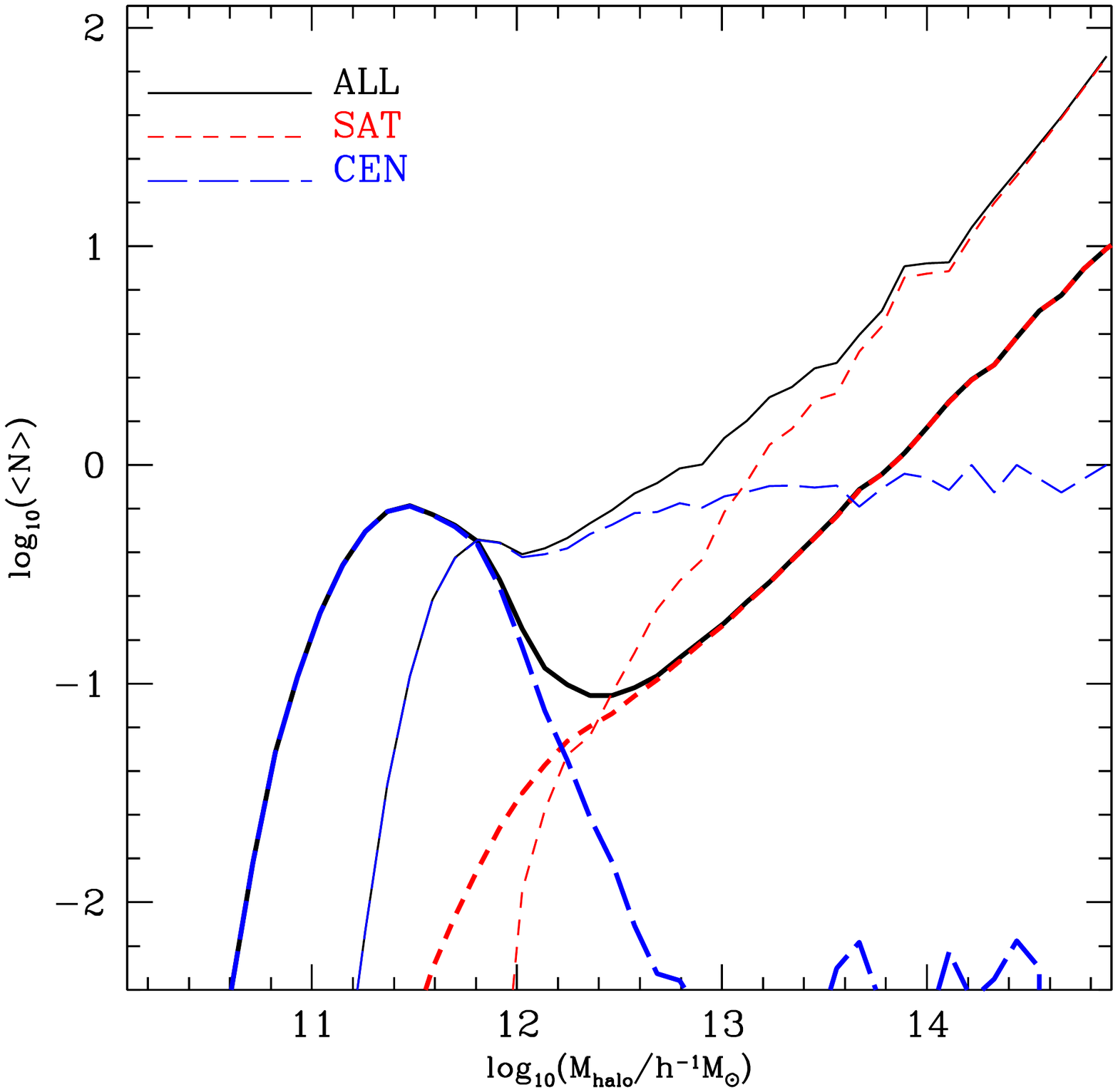}
  \caption{Predicted spatial clustering of HI galaxies. The left panel shows the 
    projected 2-point correlation function of HI-rich galaxies at $z$=0, with symbols
    and errorbars indicating observational data from \citet{Meyer07}. Lines show 
    predictions from three different models, as labelled, for galaxies with HI masses 
    $M_{\rm HI}> 10^{9.25}\,h^{-2}\, M_{\odot}$, which is consistent with threshold assumed
    in the observational measurement. The right panel shows the halo occupation distribution for 
    HI-rich galaxies (heavy curves) and $r$-band selected galaxies (light curves) in 
    the \citet{Lagos11} model, using the same data as in the left panel.
    ALL, SAT and CEN refer to all, satellite and central galaxies respectively.}
  \label{CF}
\end{figure}

Both the older post-processed models and the newer self-consistent ones predict clustering
of HI-selected galaxies that are broadly consistent with observations, which indicate that the
clustering strength of HI-selected galaxies is weaker than that of their optically-selected 
counterparts \citep[cf.][]{Meyer07,Martin12}. For example, \citet{Kim10} and \citet{Kim13} 
demonstrated that the post-processed \citet{Bower06} and \citet{Font08} models\footnote{Their cold 
  gas masses were converted to HI masses using the approach of \citet{Baugh04}, which is tuned to 
  recover the global densities of HI and H$_2$ at $z$=0.} and the self-consistent \citet{Lagos11} 
model, all variants of the {\small GALFORM} model of \citet{Cole00}, make predictions that are in good
agreement with the observed 2-point correlation functions (cf. left panel of Fig.~\ref{CF}, with data
points drawn from \citealt{Meyer07}). The models suggest that it is the central galaxies of halos 
with masses $\approx 10^{11}h^{-1}$M$_{\odot}$, which are more likely to be HI-rich, that determine 
the shape of the 2-point correlation function. 

Interestingly, the \citet{Font08} model, which seeks to capture the effects of ram-pressure 
stripping on the ISM of satellites and the consequences of this for their observed colours, 
predicts stronger clustering on small scales than the \citet{Bower06} and \citet{Lagos11} models, 
which provide a better description of the observed data points. 
This suggests that treating the ISM of a galaxy either explicitly in its HI and H$_2$
phases or simply as a monolithic cold gas phase has little effect on predicted 
clustering. Physically we expect how the ISM is modelled to influence how it is stripped -- 
lower column density HI should be stripped preferentially while higher column density H$_2$ should 
resist stripping, which is what is found in hydrodynamical simulations 
\citep[e.g.][]{Quilis00,Bekki14} -- although it is plausible that the behaviour of a single phase
cold gas ISM and a multiphase ISM in which H$_2$ is retained is similar. Nevertheless, the treatment 
of the ISM in even the newer models is too coarse-grained to capture this, and ram pressure 
stripping modelling requires further development \citep[e.g.][]{Book10}.

The models also predict that the halo occupation distribution (HOD) for HI galaxies (the mean number
of galaxies with a given HI mass per host dark matter halo as a function of host halo mass; cf. 
right hand panel of Fig.~\ref{CF}, based on the \citealt{Lagos11} model) has a functional form that 
is distinct from that usually employed for optically selected galaxies \citep[e.g.][]{Wyithe10}. We
show Fig.~\ref{CF} the HOD for HI-selected galaxies (heavy solid curve) split into centrals (heavy 
long-dashed curve) and satellites (heavy short-dashed curve); for comparison we show also the HOD 
for optically-selected ($r$-band) galaxies (light solid curve), with light long- and short-dashed 
curves indicating centrals and satellites respectively. The contribution to the HI-selected HOD from 
centrals is negligible at higher halo masses ($\gtrsim 10^{13} h^{-1} \rm M_{\odot}$), whereas the 
measured HOD in the $r$-band is consistent with the usual functional form assumed in the literature.
Reconstructing the HOD of HI-selected galaxies will provide an important test of the treatment of 
ram pressure stripping of the models.

\section{Probing Galaxy Kinematics with HI}
\label{sec:kinematics}

\begin{figure}[h]
\centering
  \includegraphics[width=0.3\textwidth]{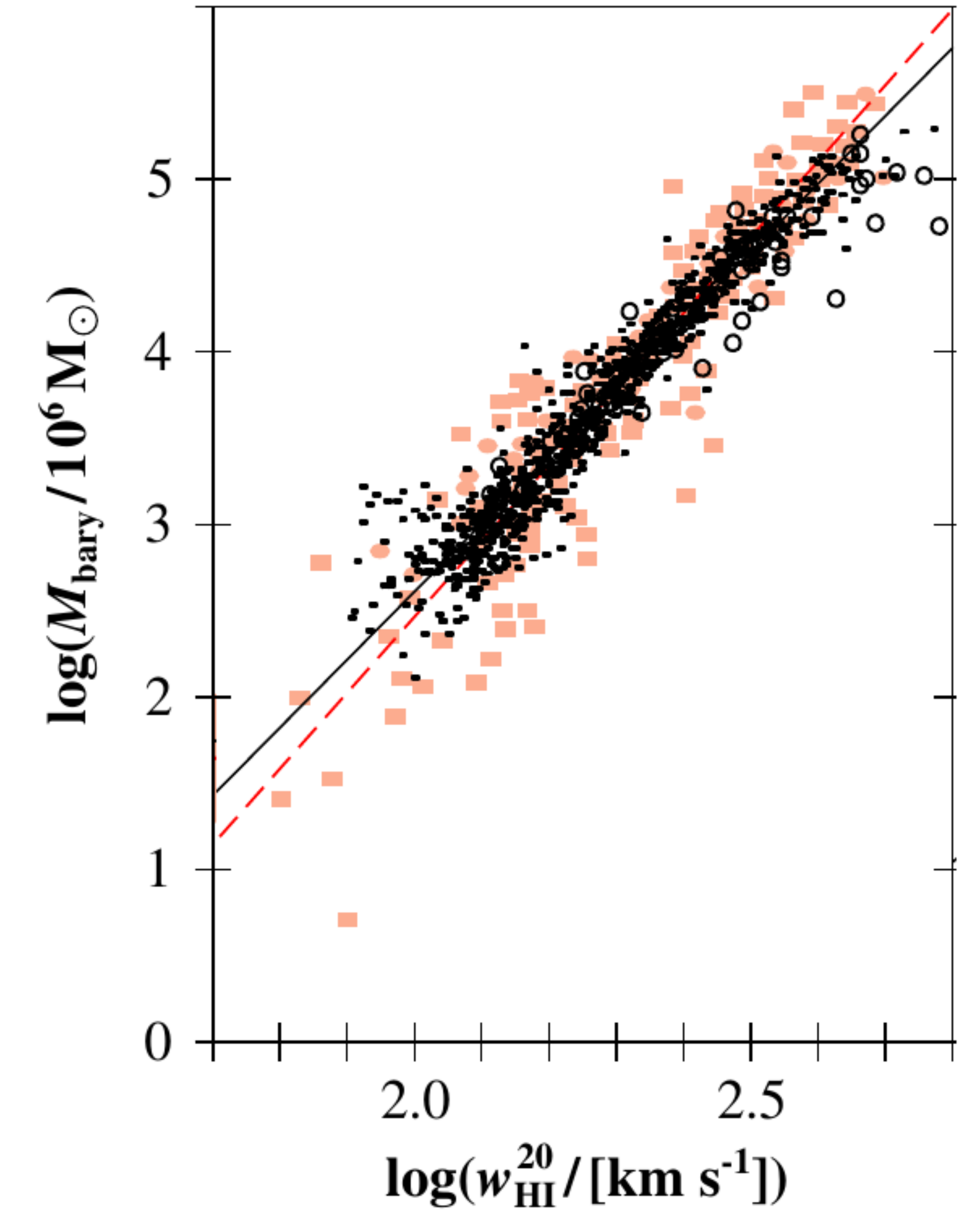}
  \includegraphics[width=0.53\textwidth]{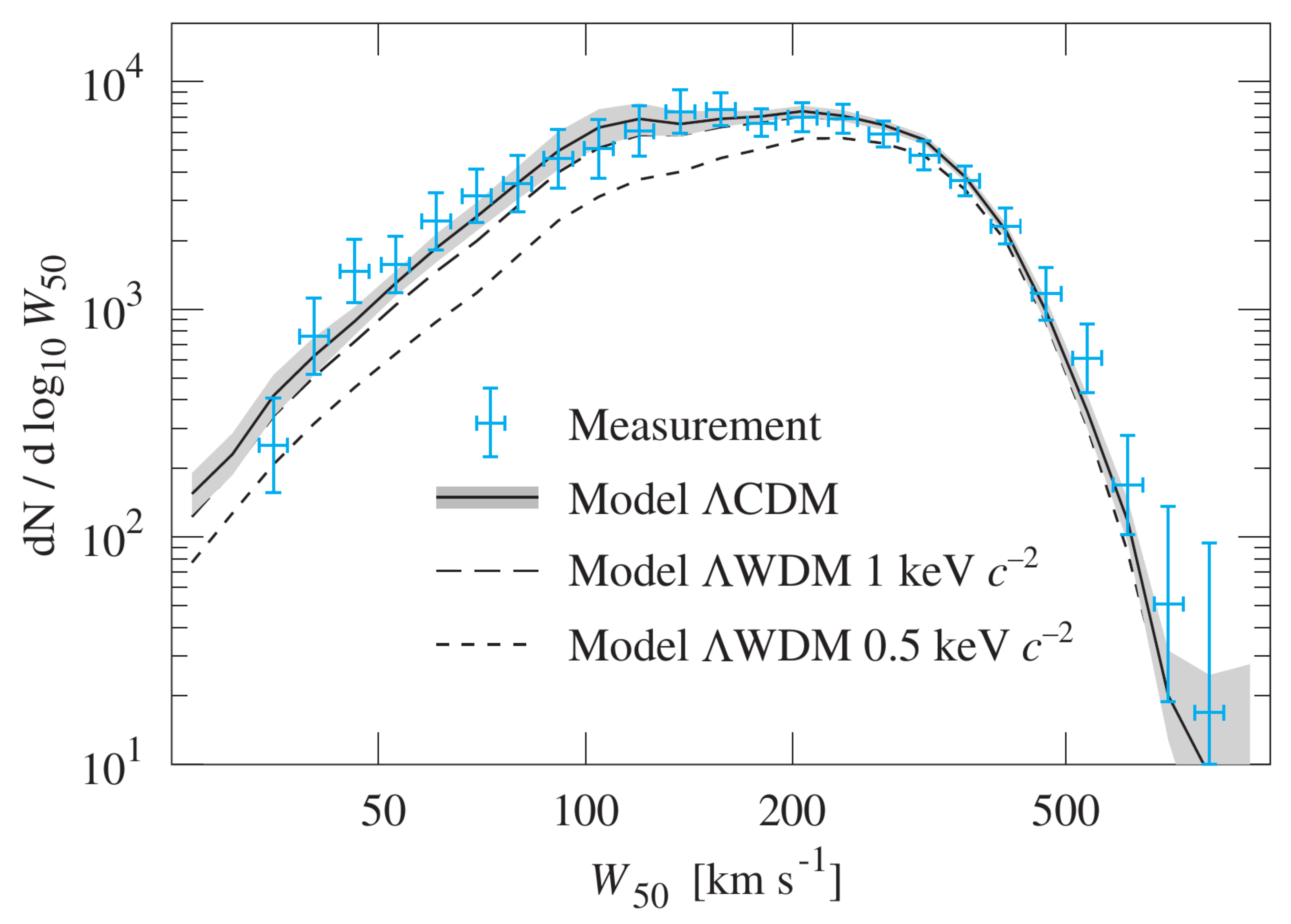}
  \caption{The left panel shows the baryonic mass as a function of HI velocity width measured 
at $0.2$ of peak flux, $w_{\rm 20}$. Simulated spirals and elliptical galaxies from 
\citet{Obreschkow09} are shown as filled and empty circles, respectively, while observations 
from \citet{McGaugh00} are shown as solid squares. The
    right panel shows the distribution of HI velocity widths measured at $0.5$ of peak flux, $w_{\rm 50}$ 
for the simulation of \citet{Obreschkow09}. Observations from \citet{Zwaan10} are shown as symbols.}
  \label{TFrel}
\end{figure}

The baryonic content of galaxies in the local Universe can be probed using the Tully-Fisher 
relation between galaxy velocity width measured by HI in emission and baryonic mass 
(see left panel of Fig.~\ref{TFrel}). The dramatic increase in the number of galaxies with good 
quality HI emission lines measured by the SKA will allow us to test our understanding of galaxy 
kinematics 
for statistical samples of galaxies as a function of epoch with the most stringent 
data ever available. Semi-analytical models have proven to be particularly useful in modelling 
HI-derived kinematics, with work presented in \citet{Obreschkow09}, 
\citet{Obreschkow09b} and most recently \citet{Obreschkow13} making it 
possible to build catalogues including complete HI emission lines (with velocity width, 
peak flux, 
and integrated flux) for millions of galaxies in a semi-analytic model
derived from the Millennium-I Simulation \citep[cf.][]{Obreschkow09}.

Using this catalogue, the consistency of the galaxy velocity function (i.e. comoving number 
density of galaxies with a given rotational velocity) predicted in a Cold Dark Matter (CDM) 
universe with observations has been assessed.\citet{Obreschkow13} replicated selection effects 
and completeness of observed galaxies to show that the reported discrepancy between the 
predicted and observed velocity functions claimed in the literature \citep[e.g.][]{Zwaan10} 
is more apparent than real, arising from observational incompleteness and erroneous corrections 
to the observed data to recover galaxy velocities. The right panel of Fig.~\ref{TFrel} shows 
the distribution of HI velocity widths predicted by the model of \citet{Obreschkow09} after 
taking into account sky coverage, completeness and blending of sources, compared to observations. 
The excellent agreement suggests that comparison of predicted and observed velocity functions 
must be to directly model HI emission lines using predicted galaxy properties and to replicate 
observations using model data.

The ability to study evolution in the internal kinematics of galaxies offer 
new and powerful insights into a property of galaxies that has traditionally
been difficult to study but which plays a crucial role in shaping not only a
galaxy's morphology but also the distribution of gas, the impact of feedback,
the growth of its supermassive black hole, etc... namely angular momentum.
This will be an area in which  the SKA can have a transformational impact.

\section{Future Developments}
\label{Future}

We now survey anticipated developments in the models, focussing on three key areas. First, 
the distribution of gas in the circum- and inter-galactic media (CGM and IGM respectively);
second, feedback within galaxies; and third next generation synthetic surveys. 

\paragraph{Distribution of Gas in the CGM and IGM.} Large HI 21 cm emission line 
surveys of local galaxies \citep{Zwaan05} and absorption-line measurements in the 
spectra of quasi-stellar objects \citep{Noterdaeme12} have allowed accurate 
measurement of the evolution of the global density of atomic hydrogen over the
redshift range $0 \leq z \lesssim 4$, and these suggest little evolution of the 
global density of atomic hydrogen, $\Omega_{\rm HI}$, with time. The latest semi-analytics 
models account for HI in galaxies and can account for evolution of $\Omega_{\rm HI}$ over the 
range $0 \leq z \lesssim 2.5$, but 
predict lower values than observed at higher redshifts by a factor of $\sim$ 2, suggesting
a large reservoir of neutral hydrogen outside of galaxies \citep{Lagos11}. This 
neutral gas does not need to be outside of the galaxy's halo, but instead it could be in the 
CGM. Indeed, observations indicate that it is 
within $\approx$ 20-40 kpc of the galaxy \citep{Krogager12}. 

The interpretation of the results in \citet{Lagos12} are consistent with the 
results of recent hydro-dynamical simulations \citep{VandeVoort12,Dave13,Rahmati14}. 
The challenge for semi-analytical models is to account for gas in the CGM and IGM in a 
self-consistent way. One approach will be to couple semi-analytical models
to a hydrodynamical description of the IGM. Such a technique has been recently 
applied by \citet{Moster14} to idealised galaxy mergers. On larger scales, work 
is already underway (by some of the authors) to characterise the sensitivity of 
the IGM to the choice of cooling and feedback prescriptions adopted in 
hydrodynamical simulations, to develop a simple predictive model for IGM properties 
based on the underlying dark matter density field. A complementary approach is to
use state-of-the-art gas dynamical simulations to study the distribution of neutral 
gas in haloes combined with semi-analytical calculations of the neutral gas 
distribution at cosmological reionization \citep{Kim13b}. Both techniques will lead 
to an ultimate answer to what is the predicted density of HI all the way from $z = 0$ 
to $z = 12$ including all possible sources of neutral gas in an ab-initio galaxy 
formation simulation. A natural by-product of this work will be predictions for HI 
absorption.

\paragraph{Impact of Stellar Feedback.} Feedback, the process by which stars and 
accreting compact objects interact with and modify their environment, is what makes
galaxy formation modelling particularly challenging. In particular, feedback from 
supernovae (SNe) represents a long standing problem. Traditionally it has been
parametrized to reproduce the faint-end of the luminosity function \citep{Guo11}, but
ignored physical properties of the galaxy such as the density of the ISM or the amount
of energy released. Recent models \citep{Creasey12,Hopkins12,Lagos13} now incorporate a 
treatment of SNe-driven outflows that account for the physical properties of the galaxy,
and have, for example, revealed the importance of gas surface density in regulating
mass entrained in the outflows, which was ignored in previous models. 

This improved description gives predictions of mass entrainment and outflow velocities 
in good agreement with observations. However, the development of a galaxy-wide outflow does 
not necessarily imply long-term disruption of the galaxy's gas reservoir, and
improvements are necessary to calculate the evolution of such outflows -- when are they
confined by the galaxy's halo, and when do they escape? These question also relate
to the condition for gas expelled from the galaxy to be reincorporated into it.
Whether or not gas that is expelled from the disc returns
to the galaxy or is lost to the gaseous halo is one of the key unknowns in galaxy 
formation, which lead to very different star formation histories of galaxies. For 
example, \citet{Henriques13} showed that a different parametric form for 
reincorportation will help the build-up of the stellar mass predicted by models to 
get closer that what is inferred from observations. Resolving this problem
will require not only extensions of the models but careful comparison with 
gas-dynamical simulations of galaxies in both a controlled non-cosmological setting
and in cosmological simulations. 

\paragraph{Next Generation Synthetic Surveys.} The importance of the models in the design 
and astrophysical interpretation of future surveys means that the creation of synthetic 
surveys is central to future developments of the models. We focus on two aspects.
First, \emph{radio continuum emission}. The SKA is predicted to detect hot, diffuse gas 
in the cosmic web, undetectable with current instruments, through its emission in the radio 
continuum. At the same time, these surveys will detect radio continuum emission from 
galaxies, driven by star formation and AGN. Some of the framework to calculate radio 
continuum emission is already in place, e.g. the work of \citet{Fanidakis11} on the 
mass and spin growth of black hole engines of AGN. However,
further development is required to account for emission from star forming regions and the
cosmic web, involving e.g. hybrid hydrodynamical and semi-analytical modelling.
Second, \emph{molecular line emission}. The SKA will survey molecular emission from 
the high-redshift Universe \citep{Geach12} using the novel technique of intensity 
mapping \citep{Carilli11}. 
This gives two-dimensional spatial information about the emission line at a given 
redshift; imaging is obtained by aggregating emission from thousands of galaxies on 
very large scales to produce the summed signal of galaxies that are not individually 
detected. Future models must account for e.g. thermalisation of the ISM of galaxies 
by the cosmic microwave background and e.g. the full spectral line energy distribution 
of the most used molecular emission line, carbon monoxide (CO).


\section{Testing Dark Matter Models with the SKA}
\label{sec:darkmatter}
\begin{figure}[h]
\centering
  \includegraphics[width=0.44\textwidth, trim=100 0 100 0, clip]{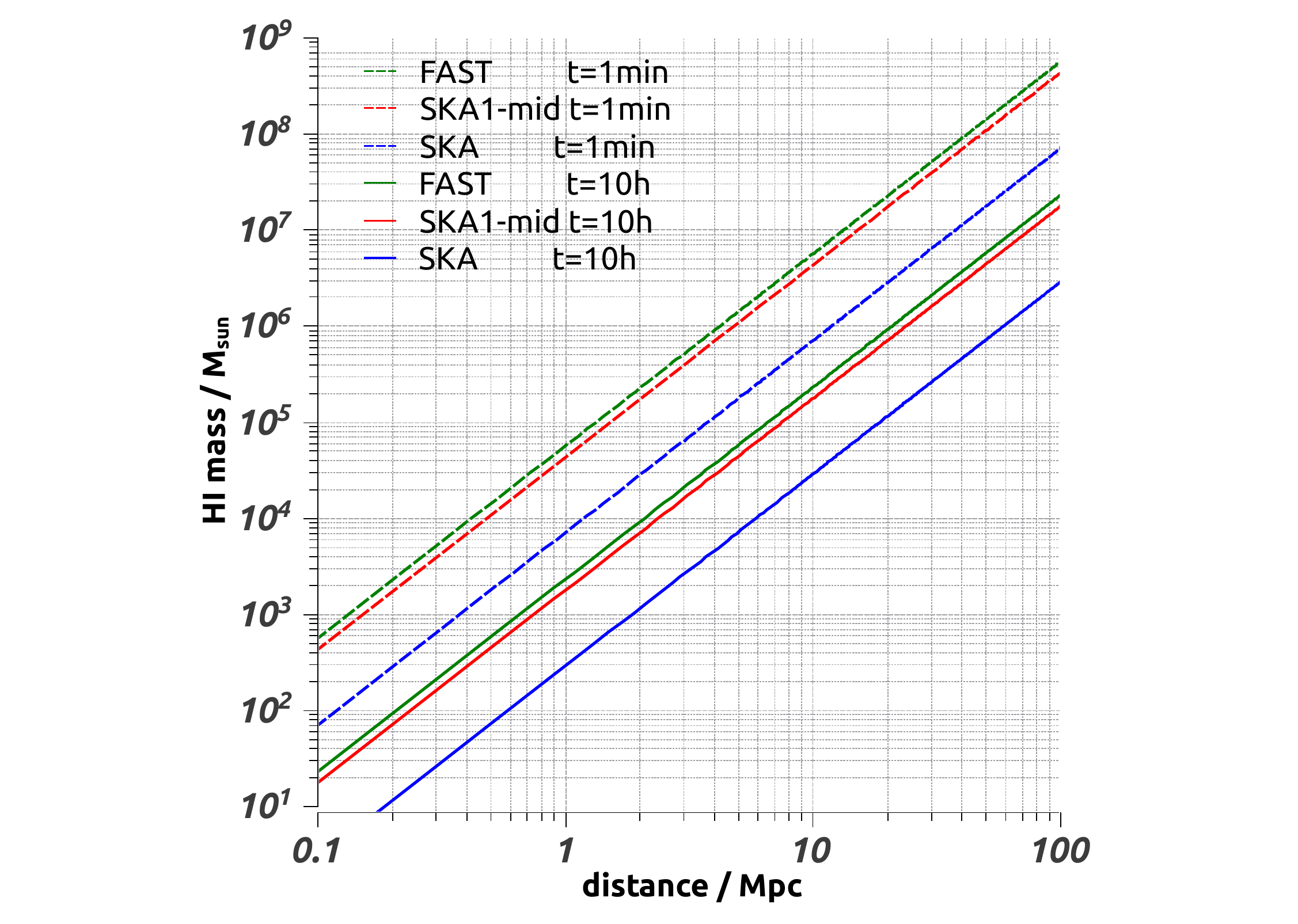}
  \includegraphics[width=0.44\textwidth, trim=100 0 100 60, clip]{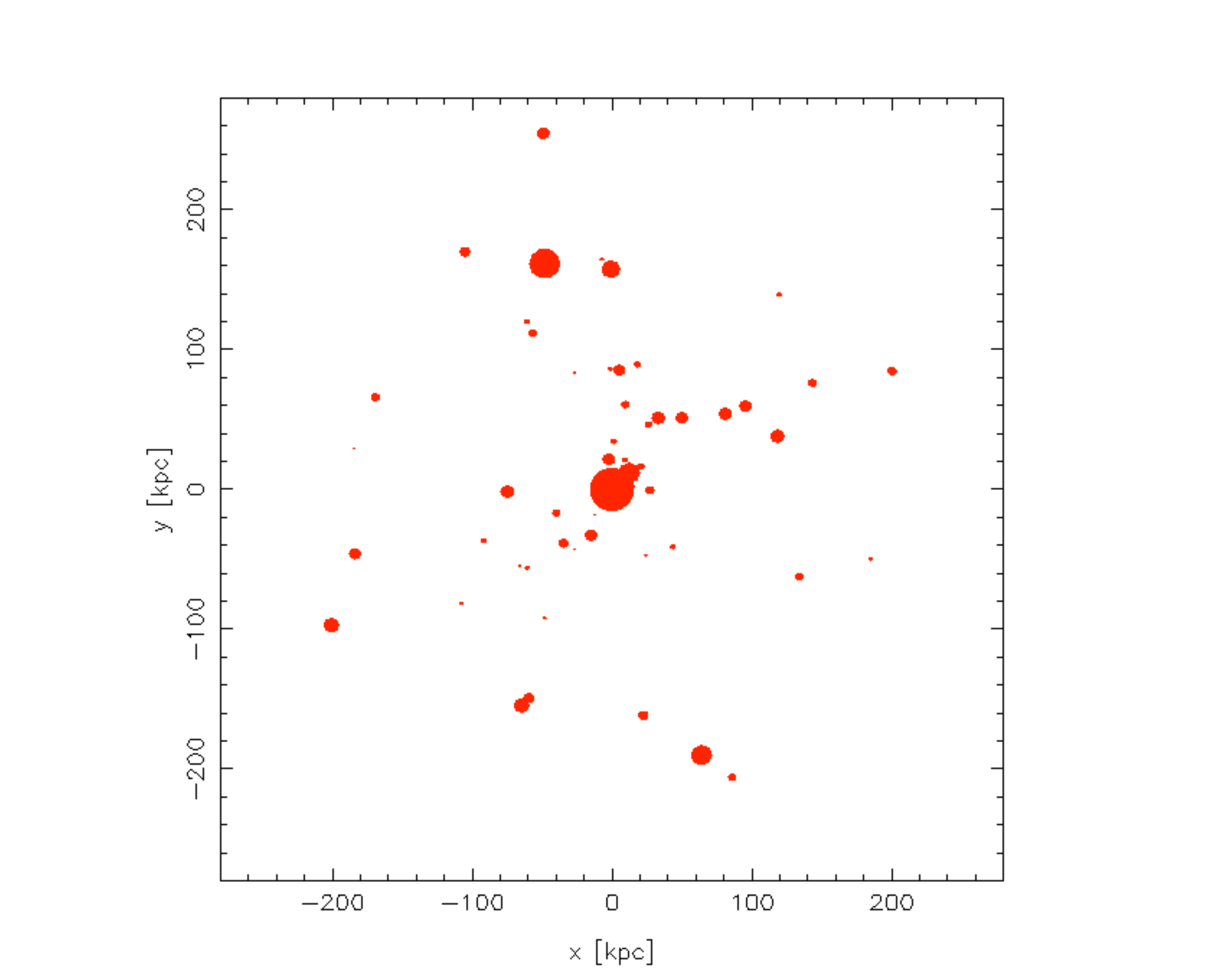}
  \caption{{\emph {Left Panel}}: HI detection limits of FAST (green), SKA1-MID (red), 
    and SKA2 (blue). Dashed and solid lines show 1min and 10h integrations. 
    {\emph {Right Panel}}: A preliminary result of the HI distributions of a 
    Milky Way-like galaxy from the Aquarius simulation. The HI mass limit is 
    $10^4 \rm M_{\odot}$.}
  \label{fig:dark_matter}
\end{figure}

The CDM model for cosmological structure formation is well established.
However, accumulating evidence during the past $15$ years has revealed that 
its predictions are at odds with observations on galactic and sub-galactic scales. 
Numerical simulations of the CDM model predict abundances of substructure 
halos (subhalos) that are an order of magnitude larger than the numbers of satellite 
galaxies observed around nearby galaxies. This has led to modifications of CDM to be
proposed, ranging from coupling between CDM and radiation \citep{boehm.etal.2014} 
to self-interacting dark matter \citep{Spergel00,Qin01} to Warm Dark Matter 
\citep[WDM; cf.][]{Bode01}, all of which suppress the abundance of 
subhalos. For example, \citet{lovell.etal.2014} have shown that the subhalo abundance
in a WDM galaxy-mass halo at $\rm M_{\rm sub} > 10^7 \rm M_{\odot}$ is approximately a factor of 
10 smaller than in the corresponding CDM halo, and \citet{kennedy.etal.2014} have shown 
that the magnitude of the effect depends strongly on WDM particle mass. 

HI observations may reveal optically invisible satellite galaxies around nearby galaxies. 
For example, the Arecibo ALFALFA HI survey of the local Universe \citep{Haynes11} has found more 
than $15,000$ extragalactic HI sources within $\sim$250~Mpc, with the lowest HI mass 
of a few $10^6 \rm M_{\odot}$. SKA1-MID will have higher sensitivity and a much larger field of 
view (FoV, 0.7 deg $\times$ 0.7 deg) than Arecibo; SKA2 will $10$ times higher sensitivity and 
$20$ times larger FoV. 
The left panel of Fig.~\ref{fig:dark_matter} shows the HI detection limits 
for the planned 500-m FAST \citep{Nan11}, SKA1-MID, and SKA2, at $1$ min and $10$ hours 
integrations, with velocity linewidth of $200$~km/s and signal-to-noise of $6$. 
SKA1-MID (SKA2) will be able to detect HI masses of a few $10^6 \rm M_{\odot}$ ($10^5 \rm M_{\odot}$) 
respectively at a distance of $50$Mpc and with a 10-hour integration; this corresponds to host 
halo masses of order $\sim 10^7 - 10^8 \rm M_{\odot}$, assuming that these halos contain a few percent 
of HI mass. This implies that the SKA can potentially discriminate
between the CDM and WDM models by detecting the difference in predicted subhalo abundance.

The right panel of Fig.~\ref{fig:dark_matter} shows the predicted distribution of HI gas in 
satellites in a Milky Way-type halo, taken from the Aquarius Simulation \citep{Springel08} 
coupled to the semi-analytical model of \citet{Fu10}, with an HI cutoff mass of
$10^4 \rm M_{\odot}$; the influence of reionization is accounted for, and dot size is proportional 
to HI mass. This indicates that the optically invisible satellite haloes of nearby galaxies may 
be visible by the SKA, if they contain enough HI gas. By surveying $\sim$
100 nearby galaxies within a few tens Mpc, and comparing with the results of numerical 
simulations such as the Aquarius Simulation, the SKA may be able to provide independent
limits on dark matter particle candidates and, in doing so, guide experimental astro-particle
physicists in their efforts to detect the dark matter particle directly in the lab.

\vspace{-0.4cm}\section{Summary}

Neutral hydrogen plays a fundamental role in galaxy formation, as the raw material
from which stars form and galaxies are built, yet much of what we know about the
HI and H$_2$ content of galaxies derives from the local Universe. The SKA will change
this dramatically, revolutionising our understanding of neutral hydrogen in galaxies 
over cosmic time and its role in galaxy formation and evolution. Recognition of this
has led to rapid developments in theoretical galaxy formation models over the last
$\sim$ 5 years, which we have reviewed. Key has been a radical rewriting of the way
star formation is modelled \citep{Lagos10}, linking explicitly a galaxy's star 
formation rate to its H$_2$ abundance, in agreement with both high resolution
observations of star-forming regions in galaxies (e.g. \citealt{Bigiel08}) and the
results of detailed numerical simulations (e.g. \citealt{Glover12}). 

This has resulted
in pleasing consistency between model predictions (e.g. \citealt{Lagos11,Kim10}) and 
observations of HI in galaxies in the local Universe 
(e.g. \citealt{Zwaan05,Meyer07,Martin10}), as well as providing insights into how gas 
is distributed in the high redshift Universe (e.g. \citealt{Lagos12}). At the same time
the models have indicated the kind of observations that are necessary to better
understand the physical processes that drive galaxy formation, most notably feedback
\citep{Kim13}, and to constrain dark matter particle physics by placing limits on 
the abundance of low-mass dark matter halos containing cold gas.

The models are still evolving, however, and we have highlighted key areas where 
further developments, including the development of hybrid schemes in which 
semi-analytics and gas dynamical simulations are coupled to better capture the  
distribution of gas outside of galaxies and how this gas accretes onto galaxies; 
the modelling of feedback from supernovae; and more realistic synthetic galaxy 
surveys. These are topics where the authors will be focussing their collective
efforts in the coming years.

\bibliographystyle{mn2e_trunc3}
\bibliography{galform_AASKA14}

\end{document}